\documentclass[journal]{IEEEtran}
\usepackage{blindtext, graphicx, algorithm2e, mathtools, amsmath}

\usepackage{cite}

\DeclarePairedDelimiter{\norm}{\lVert}{\rVert}
\DeclareMathOperator*{\argmin}{argmin}
\DeclareMathOperator*{\argmax}{argmax}

\usepackage[english]{babel}
\usepackage{blindtext}
\usepackage{draftwatermark}
\SetWatermarkText{PREPRINT}
\SetWatermarkScale{1}
\ifCLASSINFOpdf
\else
\fi

\begin{document}
\pagenumbering{gobble}

%
\title{Extracting Blink Rate Variability from EEG Signals}



%
\author{\IEEEauthorblockN{Rafal Paprocki,
Temesgen Gebrehiwot,
Marija Gradinscak and
Artem Lenskiy}

\IEEEauthorblockA{Korea University of Technology and Education\\
1600, Chungjeol-ro, Byeongcheon-myeon,
Dongnam-gu, Cheonan-si, Chungcheongnam-do 31253,\\ Republic of Korea\\
email: lensky@koreatech.ac.kr, rafal.paprocki@gmail.com}
}



\maketitle

\begin{abstract}
Generally, blinks are treated on equal with artifacts
and noise while analyzing EEG signals. However, blinks carry 
important information about mental processes and thus it is
important to detect blinks accurately. The aim of the presented study
is to propose a blink detection method and discuss its application for
extracting blink rate variability, a novel concept that might shed
some light on the mental processes. In this study, 14 EEG recordings
were selected for assessing the quality of the proposed blink
detection algorithm.
\end{abstract}


%
\IEEEpeerreviewmaketitle

\section{Introduction}
Blinking is a semiautonomic closing of the eye lids.
Blinks keep eyes protected against potentially damaging
stimuli, such as bright lights and foreign bodies like dust.
The sudden changes in an image due to saccades or blinks
does not interfere with our subjective experience of continuity
\cite{ref1}. The very act of blinking suppresses activity in several
areas of the brain responsible for detecting environmental
changes, so that one experiences the world as continuous.
Researchers have shown synchronous behavior in
blinking between listener and speaker in face-to-face
conversation \cite{ref2}. Reduced blink rate causes eye redness and
dryness also known as Dry Eye, which is a major
symptom of the Computer Vision Syndrome \cite{ref3}.
Blinks have been known to be linked to interior brain
activities. Increasing the accuracy of blink detection is of
high importance as humans look for easier methods of
collecting internal brain activity information. The detection
of the eye blinks had a huge impact in various fields. In some
Brain Computer Interface (BCI)  researchers analyzed eye blinks to determine the pattern and the duration between blinks. After collecting this analysis, the information was used with a device that could control a computer similarly to how we use a computer mouse. This implementation of the
use of blinks has opened a wide door to new possibilities for
disabled people \cite{ref4}. Another area where blinks play an
important role is in the prevention of car accidents. The World Health
Organization (WHO) has announced that the ninth cause of
death, globally, is car accidents. The National Motor Vehicle
Crash Causation Survey (NMVCCS) has found that 30 percent of
car accidents happen due to the drowsiness of drivers
\cite{ref5}. It is noted that workload increases blink rate and blink
rate is known to decrease in monotonous and drowsy
conditions \cite{ref6}. Blink rate (BR) is inversely correlated with
the increase of workload so blinks can be used to detect
drowsiness before it causes damage \cite{ref6}. Researchers have
shown that blinks can play a significant role in detecting
many different brain disorders and brain activities.
Spontaneous BR has been studied in many neurological
diseases like Parkinson's disease and Tourette syndrome
\cite{ref7},\cite{ref8},\cite{ref9}. The use of blink detection does not stop there.
 Blinks are regarded as a non-invasive peripheral markers of
central dopamine activity which makes their accurate
detection more important \cite{ref10},\cite{ref11},\cite{ref12},\cite{ref13},\cite{ref14},\cite{ref15}. Researchers have studied
the synchronicity of the eye blinks in audiences, who
experienced the same method of storytelling. The eye
blink synchronization among audiences is driven by attention
cycles, which are in turn driven by emotional processing
\cite{ref16},\cite{ref17},\cite{ref18}.
Blinks are not always the most desired signals when it
comes to non-invasive brain signal measuring as many
electroencephalographs (EEG) remove them to acquire brain
data. Eye blink is one of the main artifacts in the EEG signals
\cite{ref19}. Researchers are focusing on removing these parts of the
signals to obtain clean brain signal values.
To analyze blinks and variation of inter-blink intervals it
is important to detect blinks accurately. We propose to apply
the blink detection algorithm for extracting the inter-blink
intervals that we coin the blink rate variability (BRV) in
analogy to heart rate variability. We construct BRV for
subjects taking memory tests. We further compare the
numbers of detected blinks by the algorithm and by manual
counting.

\section{Experimental setup}

\subsection{Data Acquisition}
\begin{figure}[!htbp]
\includegraphics[width=3.5in]{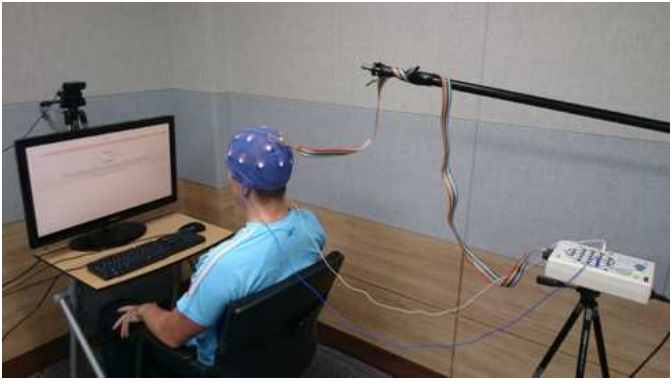}
\caption{{\bf Experimental setup}}
\label{fig1}
\end{figure}

The video stream was captured with a Pointgrey Flea3 USB
camera. Video stream was stored on a disk drive to be
processed in the future. Simultaneously, EEG signals were
recorded. For the recording of EEG signals, we employed
a Mitsar-EEG 201 amplifier and used WinEEG
software. The electrodes were placed according to the
international “10-20 system” \cite{ref20}. Electro-gel was
injected into electrodes' hollow in order to decrease the
electrode-skin resistance. Currently, the EEG signals were
recorded for the purpose of eye blink detection. In the future,
we are planning to analyze EEG to detect various types of
brain activity. The experimental setup is shown in fig.\ref{fig1}.

\subsection{Testing Procedures}
The testing software was developed using Java in such a way that it does not require any interventions. The procedure consisted of a five minute reading session and a five minute memory test session. Before the memory test, a passage about Ethiopia is given. After reading the passage, users are presented questions one by one. In this paper, we focus on detecting blinks while subjects are reading the passage and answering questions about the passage. Overall, 28 people participated in the experiment. Among them 14 subjects were dropped due to falling asleep, adjusting the cap or constant head movements that resulted in significant noise.

\section{Eye blink detection procedure}
\begin{figure}[!htbp]
\includegraphics[width=3.5in]{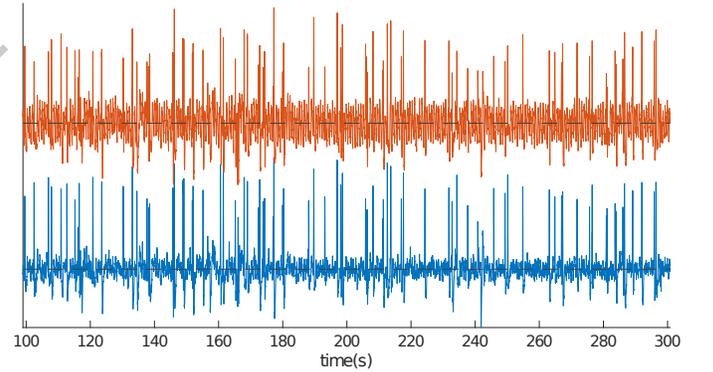}
\caption{{\bf Original Fp1-Fp3 and Fp2-Fp4 electrode pairs}}
\centering
\label{fig2}
\end{figure}

Electrodes are applied to the head according to the 10-20
system. We used bipolar montage, which means we
determined the potential between Fp1 and Fp3, along with Fp2 and
Fp4. Fig. \ref{fig2} presents EEG signals for both pairs.
EEG signals were recorded while participants were taking
the tests and imported in form of CSV files to Matlab for
further analysis. The process of blink detection can be
divided into two stages: preprocessing and blink detection.
The preprocessing stage consists of the following steps: (a) normalization and
bandpass filtering (b) cutting of extreme amplitudes using the 
estimated cumulative distribution function that characterizes
the signal amplitude's distribution, (c) independent
component analysis, and (d) selection of the component with eye blinks. The blink detection stage
consists of (e) signal thresholding, (f) candidate extraction,
and (g) polynomial fitting with finding maximum in the
polynomial function. Both stages, preprocessing and blink detection, have been presented in listings algorithm 1 and algorithm 2 respectively.

\begin{algorithm}[]

\textbf{input}: $\mathbf{X}$ is $2 \times N$ fp1-fp3 and fp2-fp4 signals\\
\textbf{output}: $y$ is the blink component\\

$H$ is the Heaviside function\\
$filt(\mathbf{X},f_l, f_h)$ is a sub-band filter, where $f_l$ and $f_h$ are low and high cut off frequencies\\
\emph{$CDF(x)$} estimates the cumulative distribution function for $x$\\
\BlankLine

\SetAlgoLined

\For{$k = 1$  to  $2$}{
    $t = 6 \cdot \sigma(\mathbf{X}_{k})$ \\
     \If{ $skewness(\mathbf{X}_{k}) \geq t$ }{
        $\mathbf{X}_{k,\{i \rvert \mathbf{X}_{k} \leq t\}} = t$\\
     }

     $\mathbf{X}_{k}  = filt(\mathbf{X}_{k},f_l,f_h)$\\
     $\mathbf{X}_{k} = \frac{\mathbf{X}_{k} - E[\mathbf{X}_{k}]}{\sigma(\mathbf{X}_{k})}$\\
    
     $\hat{F}_{\mathbf{X}_{k}} = CDF(\mathbf{X}_{k})$ \\
     $r = max(\mathbf{X}_{k}) - min(\mathbf{X}_{k})$ \\
     $\mathbf{X}_{k,\{i|\mathbf{X}_{k} \geq x_{max}\}} = min(\mathbf{X}_{k}) + r \cdot {F}_{\mathbf{X}_{k}}^{-1}(0.99)$ \\
     $\mathbf{X}_{k,\{i|\mathbf{X}_{k} \leq x_{min}\}} = min(\mathbf{X}_{k}) + r \cdot {F}_{\mathbf{X}_{k}}^{-1}(0.02)$ \\
}

\BlankLine
$\mathbf{S} = ICA(\mathbf{X})$
\BlankLine

 \eIf{ $\sum H(\lVert\mathbf{S}_1\rVert - 3 \cdot \sigma(\mathbf{S}_1)) > \sum H(|\mathbf{S}_2| - 3 \cdot \sigma (\mathbf{S}_2)$ }{
    \BlankLine
     $y = \mathbf{S}_1$ 
}{
    \BlankLine
     $y = \mathbf{S}_2$
}

\caption{The preprocessing stage}
\end{algorithm}

\begin{algorithm}[]
\label{alg:qt}

\textbf{input}: $\mathbf{y}$ - blink component,\\$f_s$ - sampling frequency
\textbf{output}: $t$ - position of blinks measured in samples

\BlankLine
\emph{$l_{min}$} - minimum arc length of a blink wave\\
\emph{$w_{min}, w_{max}$} - min/max blink duration\\
\BlankLine
\emph{$\{\mathbf{Y}, \mathbf{s}\} = segment(\mathbf{y})$} - extracts continuous segments of non-zero values from $\mathbf{y}$ and stores them as a list in $\mathbf{Y}$, $\mathbf{s}$ contains positions of segments in $\mathbf{y}$.\\
\SetAlgoLined
$l_{min} = \frac{f_s}{50}, w_{min} = \frac{f_s}{25}, w_{max} = f_s$\\
$\mathbf{y}_{\{i|\mathbf{y} < \sigma(\mathbf{y})\}} = 0$ \\
$\{\mathbf{Y}, s\} = segment(\mathbf{y})$ \\
$l = 0 $ \\
\For{$j = 1$  to  $dim(\mathbf{Y}) - 1$} {
        $ N = dim(\mathbf{Y}_j) $ \\
        $\mathbf{x} = \{i \rvert 1 \leq i \leq N\} $ \\
        $\mathbf{a} \longleftarrow \argmin\limits_{\mathbf{a}} \norm{\mathbf{Y}_j - \sum_{n=0}^3 a_n \cdot (x_0^n,x_2^n,...,x_{N-1}^n)}_2 $ \\
        $ \hat{\mathbf{y}} = (\hat{y}_0, \hat{y}_1,...,\hat{y}_{N-1})= \sum_{n=0}^3 a_n \cdot (x_0^n, x_1^n,...,x_{N-1}^n) $ \\
        $ alen = \sum_{i=0}^{N-1}  \sqrt{([\hat{y}_i] - \hat{y}_{i+1})^2 + 1/N} $ \\
     \If{ $w_{min} < N$ AND $N < w_{max}$ AND $ l_{min} < alen$ }{
        $ v = \max \mathbf{Y}_j $ \\
        $ p = \argmax\limits_{}{\mathbf{Y}_j} $ \\
        $ \alpha_1 = \frac{180}{\pi} \cdot atan2(v - \hat{y}_0,   \frac{p}{f_s})) $\\
        $ \alpha_2 = \frac{180}{\pi} \cdot atan2(v - \hat{y}_{dim(\mathbf{y})-1},   \frac{p}{f_s})) $\\
        \If{$ \alpha_1 > 80$ AND $\alpha_2 < 100$}{
            $ t_l  = \mathbf{s}_j + p - 1 $\\
            $ l=l + 1$
        }
    }
}
\caption{The blink detection stage}
\end{algorithm}

\begin{figure}[!htbp]
\includegraphics[width=3.5in]{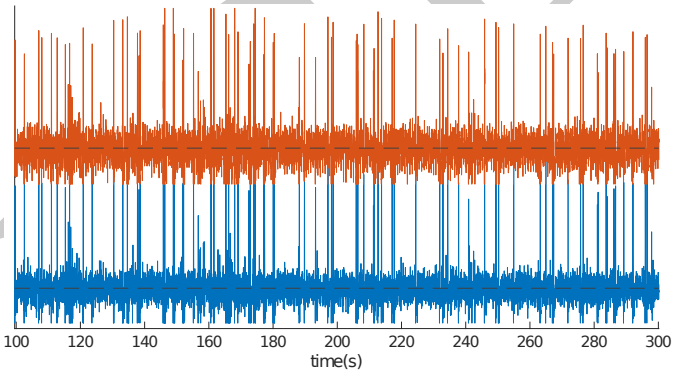}
\caption{{\bf Normalized and band-pass filtered signals}}
\label{fig3}
\end{figure}

The first step in the process of blink detection consists of cutting off very extreme amplitudes, that usually caused by touching the electrodes or cap adjustment. Right after, the signal is band-pass filtered and normalized. We applied 50th-order bandpass filter with finite impulse response. The lowest and highest normalized cut off frequencies are $f_L=0.02$ and $f_H=0.08$ correspondingly. In fig. \ref{fig3} the EEG
signals after bandpass filtering and normalization are
presented. The signals became
smoother and the lower frequency components responsible
for trends in the signals disappear after filtering. For the filtered signal an
amplitude cumulative distribution function (CDF) is
estimated. Using the CDF, we cut 2 percent of all
amplitudes from the top and 1 percent from the bottom.
The next step consists of mixing signals from two pairs
of electrodes fp1-fp3 and fp2-fp4 in such a way that led to a cleaner signal.

\begin{figure}[!htbp]
\includegraphics[width=3.5in]{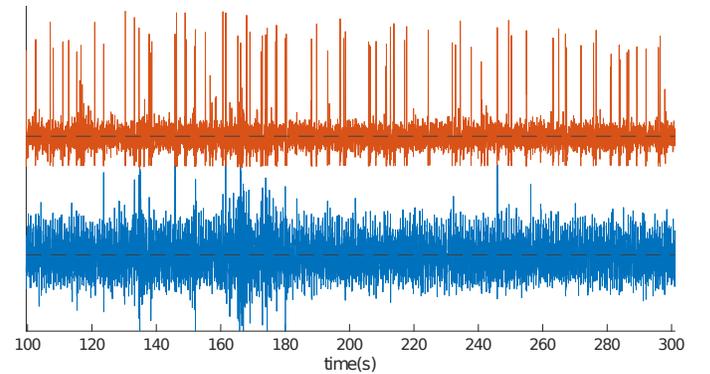}
\caption{{\bf Independent components}}
\label{fig4}
\end{figure}

\begin{figure}[!htbp]
\includegraphics[width=3.5in]{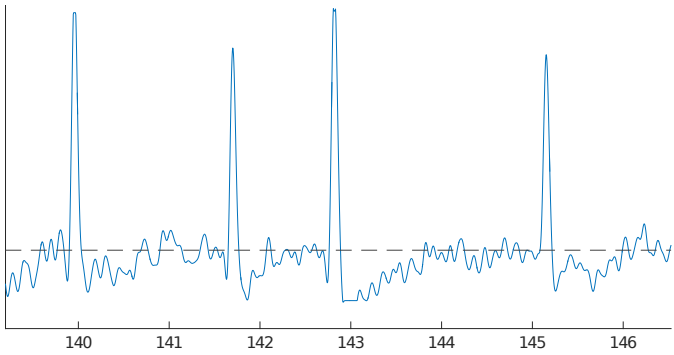}
\caption{{\bf An independent component with blinks}}
\label{fig5}
\end{figure}

\begin{figure}[!htbp]
\includegraphics[width=3.5in]{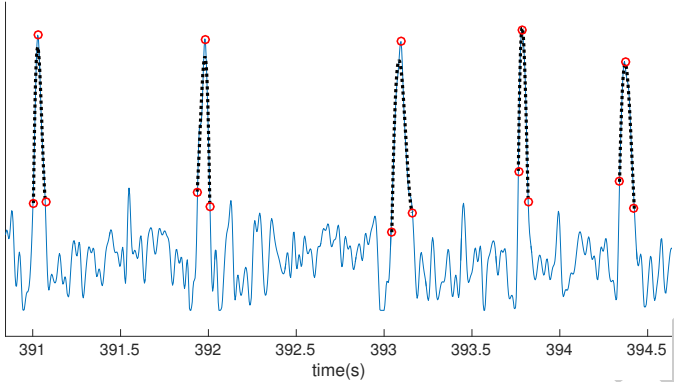}
\caption{{\bf Detected blinks}}
\label{fig6}
\end{figure}

Usually, we want to get rid of ocular artifacts from
EEG signals, as the eye blink is an artifact and leads to
interpretation problems \cite{ref21}. However, our goal is on 
contrary aims at extracting blinks from EEG. We employ
the fastICA\cite{ref22} algorithm for solving Blind Source
Separation (BSS)\cite{ref23}, which allows us to differentiate neural
activity from muscle and blinks\cite{ref24}. Independent
component analysis (ICA) consists of two steps. The
first is responsible for decorrelation or whitening, when
a correlation in the data is removed. The second stage is
responsible for the separation, which is an orthogonal
transformation of whitened signals (rotation of the joint
density). The task is to find an orthogonal matrix U
such that the projections on the orthogonal axis are
non-Gaussian \cite{ref22}. One by one, we look for the
rows of the matrix $U=(u_1,u_2)^T 	$ so a measure of
non-Gaussianity $|E(G(u^T_k x_{st}))|$ is maximized by such
$u_k$  that the length of $u_k$  is one and orthogonal to the remain row. 
The function $G$ can be any nonquadratic
function, which is twice continuously differentiable with
$G(0)=0$ . We tested a number of nonlinear
functions and the one that results in a better component
separation is the $g(z)=z^2$ skewness measure. The
$g(z)=z^2$ (skew) nonlinearity finds skew sources, but
in the case of symmetric sources is not efficient. In our
data, the skew measure shows best results due to the fact
that the waveforms corresponding to blinks are
asymmetrical (fig. \ref{fig4}).
To distinguish which of the components corresponds
to the blink component $y$, we select the component based
on the following rule
\begin{equation}
 y = 
  \begin{cases} 
   \sum H(|\mathbf{S_1}| - 3 \cdot \sigma (\mathbf{S}_1)) > \sum H(|\mathbf{S}_2| - 3 \cdot \sigma(\mathbf{S}_2)) & \text{blinks}  \\
   otherwise       & \text{}
  \end{cases}
\end{equation}
where $H$ is the Heaviside function. In (1) we count
a number of crossings at $3 \cdot \sigma(\mathbf{S}_2)$.
At the next step, we set to zero all samples that are
smaller than the standard deviation of the signal. The
samples below the threshold are zeroed. The remaining
contiguous segments are treated as blink candidates (fig. \ref{fig4}). 
Finally, a $3^{rd}$ order polynomial function is fitted to the samples within each segment. If the arc length of the
polynomial function is less than a predefined threshold,
the region is rejected (fig. \ref{fig5}). We also reject regions with
a slop of the front and the end transitions having an angle
less than 80 degrees for the front and having greater than
100 degrees for the end transition. The slop is calculated
as an angle of line connecting end points of the fitting
polynomial and its maximum (fig. \ref{fig6}).

\begin{figure}[!htbp]
\includegraphics[width=3.5in]{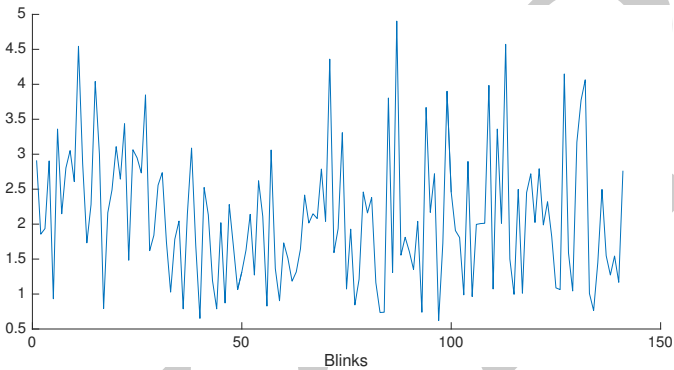}
\caption{{\bf Example of blink rate variability}}
\label{fig7}
\end{figure}

To construct blink rate variability, the time of occurrences of
consecutive blinks are subtracted. The interval between
blinks is stack-up into a series that constitutes blink rate
variability (fig. \ref{fig7}).

\section{Results and Discussion}
\begin{figure}[!htbp]
\includegraphics[width=3.5in]{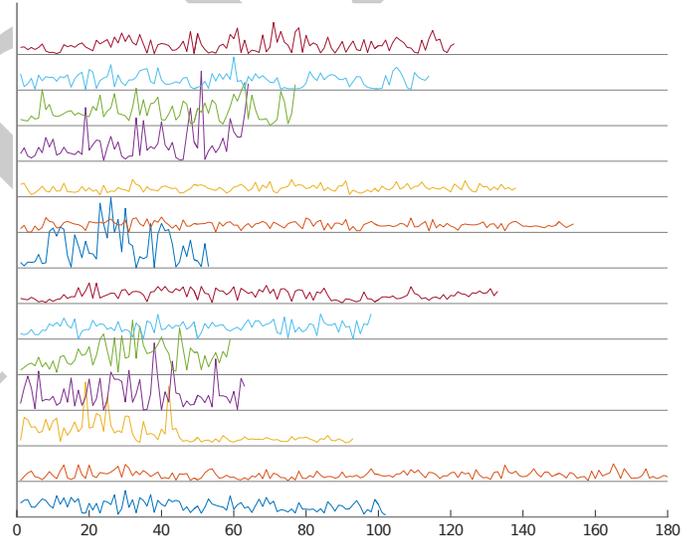}
\caption{{\bf Extracted blink rate variability for each subject during reading}}
\label{fig8}
\end{figure}

Figure \ref{fig8} demonstrates BRVs for all of the subjects during the
memory test. The abscissa is the blink interval and the ordinate
represents the length of inter-blink intervals. The BRV shows the dynamics of blink intervals extracted while memory testing for each subject.\\
To access the quality of blink detection we manually calculated the number of false positives, false negatives and true positives during the reading and the testing stages. The false positives are mistakenly detected blinks at places where blinks did not occur. The false negatives are the blinks that were missed, and the true positives are the correctly detected blinks. Based on these three categories we calculated the precision and recall characteristics that are shown in tables 1 and 2. The average precision during reading stage is $0.99 \pm 0.02$ and the average of recall is $0.99 \pm 0.01$. The average precision during memory testing stage is $0.98 \pm 0.03$ and the average of recall is $0.99 \pm 0.02$. 

\begin{table}[]
\centering
\caption{COMPARISON OF AUTOMATIC AND MANUAL BLINK DETECTION DURING READING STAGE}
\label{my-label}
\begin{tabular}{llllll}
\hline

\begin{tabular}[c]{@{}l@{}}Subject \end{tabular} & 
\begin{tabular}[c]{@{}l@{}}False\\positive\end{tabular} & 
\begin{tabular}[c]{@{}l@{}}False\\negative\end{tabular} & 
\begin{tabular}[c]{@{}l@{}}True\\positive\end{tabular} & 
\begin{tabular}[c]{@{}l@{}}Precision \end{tabular} &
\begin{tabular}[c]{@{}l@{}}Recall \end{tabular}
\\ \hline
1 &	    3 &	    0 &	100 &	0.97 &	1.00  \\
2 &	    0 &	    3 &	180 &	1.00 &	0.98  \\
3 &	    0 &	    0 &	94  &	1.00 &	1.00  \\
4 &	    0 &	    0 &	61  &	1.00 &	1.00  \\
5 &	    0 &	    0 &	60  &	1.00 &	1.00 \\
6 &	    0 &	    1 &	94  &	1.00 &	0.99 \\
7 &	    1 &	    0 &	133 &	0.99 &	1.00  \\
8 &	    2 &	    1 &	49  &	0.96 &	0.98  \\
9 &	    4 &	    1 &	148 &	0.97 &	0.99  \\
10 &	    0 &	0 &	139 &	1.00 &	1.00  \\
11 &	    3 &	0 &	62  &	0.95 &	1.00  \\
12 &	    1 &	0 &	76  &	0.99 &	1.00  \\
13 &	    2 &	3 &	113 &	0.98 &	0.97  \\ 
14 &	    0 &	1 &	122 &	1.00 &	0.99  \\ 
\hline
\end{tabular}
\end{table}

\begin{table}[]
\centering
\caption{COMPARISON OF AUTOMATIC AND MANUAL BLINK DETECTION DURING TESTING STAGE}
\label{my-label}
\begin{tabular}{llllll}
\hline

\begin{tabular}[c]{@{}l@{}}Subject \end{tabular} & 
\begin{tabular}[c]{@{}l@{}}False\\positive\end{tabular} & 
\begin{tabular}[c]{@{}l@{}}False\\negative\end{tabular} & 
\begin{tabular}[c]{@{}l@{}}True\\positive\end{tabular} & 
\begin{tabular}[c]{@{}l@{}}Precision \end{tabular} &
\begin{tabular}[c]{@{}l@{}}Recall \end{tabular}
\\ \hline
1 & 	    0 &	    1 &	187 &	1.00 &	0.99  \\
2 & 	    0 &	    3 &	212 &	1.00 &	0.99  \\
3 & 	    1 &	    1 &	76 &	0.99 &	0.99  \\
4 & 	    3 &	    3 &	63 &	0.95 &	0.95  \\
5 & 	    0 &	    2 &	146 &	1.00 &	0.99 \\
6 & 	    0 &	    0 &	124 &	1.00 &	1.00 \\
7 & 	    0 &	    0 &	144 &	1.00 &	1.00  \\
8 & 	    5 &	    0 &	67 &	0.93 &	1.00  \\
9 & 	    1 &	    1 &	197 &	0.99 &	0.99  \\
10 & 	    3 &	4 &	184 &	0.98 &	0.98  \\
11 & 	    4 &	0 &	51 &	0.93 &	1.00  \\
12 & 	    1 &	3 &	88 &	0.99 &	0.97  \\
13 & 	    0 &	8 &	181 &	1.00 &	0.96  \\ 
14 & 	    0 &	0 &	201 &	1.00 &	1.00  \\ 
\hline
\end{tabular}
\end{table}

\section{Conclusion}
Blinking is a natural, biological, semiautomatic process. It
is linked to interior brain activity and relationships between
blinks and performing tasks. The blink rate variability might
find applications in variety of fields, including car safety,
psychology or BCI.\\
In order to automatically extract blinks from large datasets of EEG
recordings, we proposed the blink detection algorithm. Basic steps such as band-pass filtering
and thresholding are applied in the algorithm. Then fastICA
is applied to find an independent component within the blinks.
The blink candidates were filtered based on the heuristics that
arc length of the blink waveform should be above a certain
threshold and the slopes of the blink waveform are within a
$[80^o,100^o]$ range.\\
Calculated the recall and precision characteristics show that the proposed algorithm is suitable for blink rate variability extraction.

%


\ifCLASSOPTIONcaptionsoff
  \newpage
\fi



%

%

 \vspace{-10 mm}
\begin{IEEEbiography}[{\includegraphics[width=1in,height=1.25in,clip,keepaspectratio]{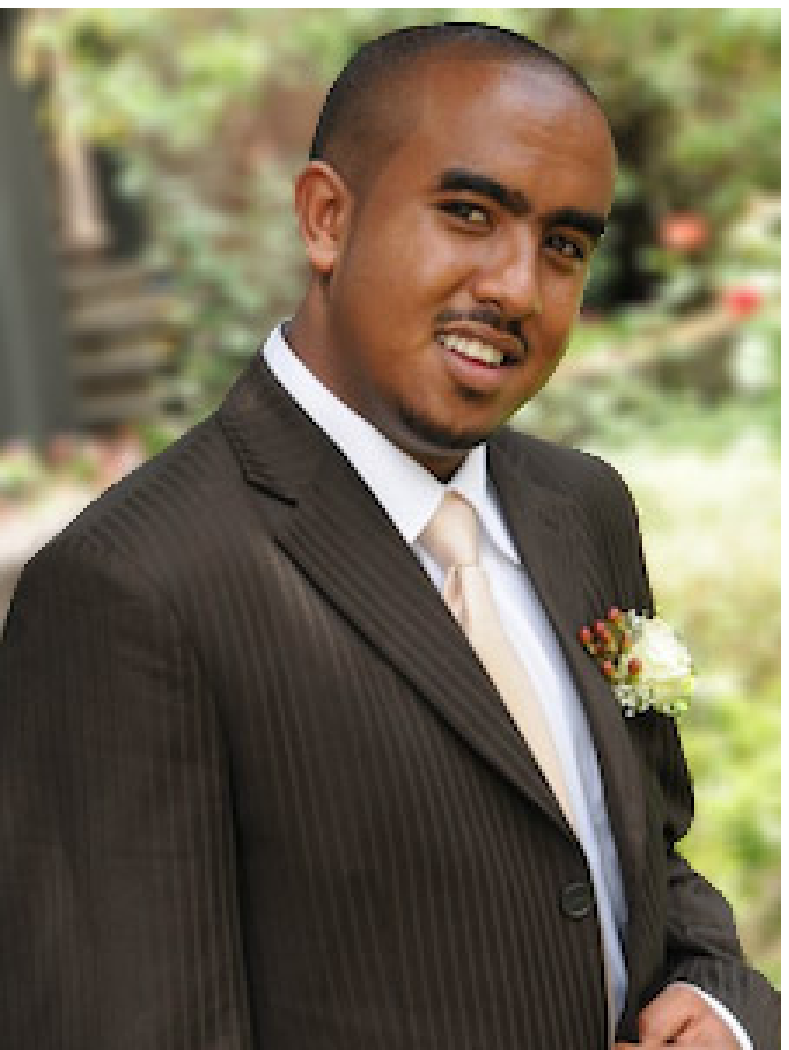}}]{Temesgen Gebrehowt}
Received first BS from Hilcoe school of computer
science science in 2011 and completing his second
bachelors in communication engineering at Korea
University of Technology and Education (South
Korea).
\end{IEEEbiography}
\vspace{-20 mm}
\begin{IEEEbiography}
    [{\includegraphics[width=1in,height=1.25in,clip,keepaspectratio]{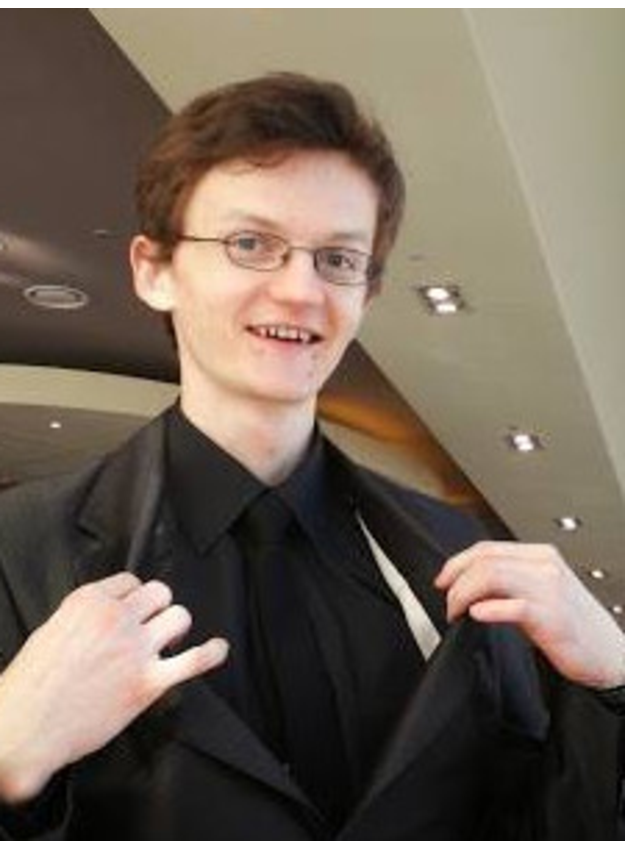}}]{Rafal Paprocki}
Completed the Bachelors and the Masters degerees in
Military University of Technology(Warsaw, Poland)
in 2010 and 2011 respectively. In 2010, Mr Paprocki
did an internship at the Università degli Studi di
Camerino (Italy). Currently he is a doctor course
student at Korea university of Technology and
Education (South Korea).
His research interest is brain signal processing with
applications to inter-blink intervals analysis.
\end{IEEEbiography}
\vspace{-20 mm}
\begin{IEEEbiography}
    [{\includegraphics[width=1in,height=1.25in,clip,keepaspectratio]{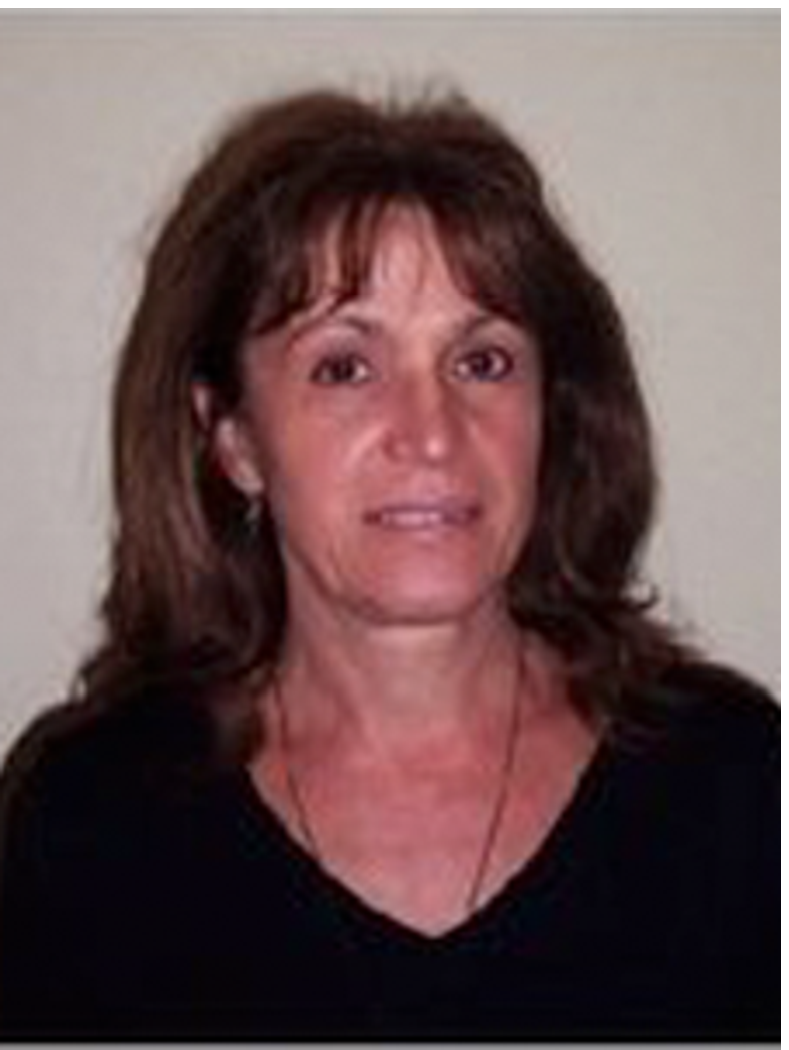}}]{Marija Gradinscak}
is an assistant professor at the Korea University of Technology and Education, South Korea. From 1996 she was actively involved in teaching and research. She received her MEng and her Ph.D from Victoria University, Melbourne.
\end{IEEEbiography}
\vspace{-20 mm}
\begin{IEEEbiography}
    [{\includegraphics[width=1in,height=1.25in,clip,keepaspectratio]{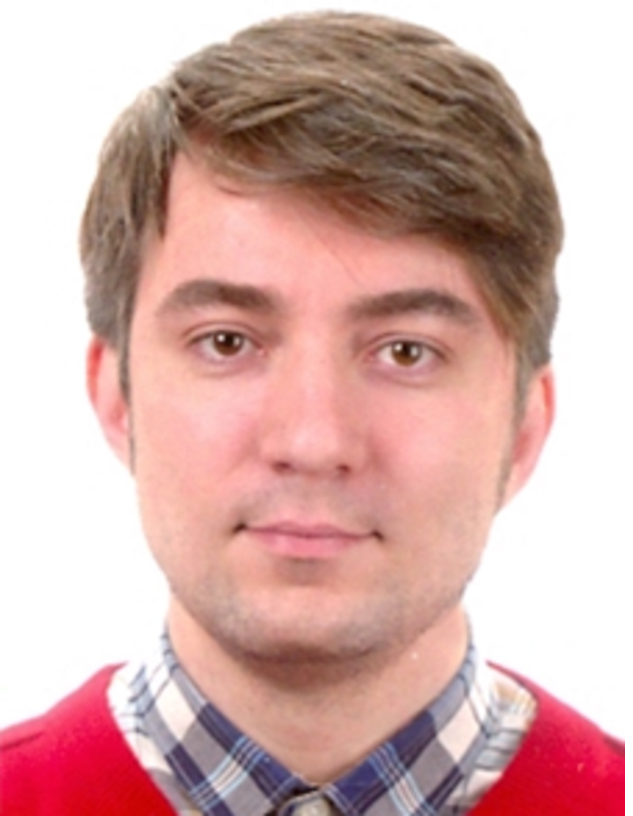}}]{Artem Lenskiy}
Received BSc and MSc degrees in computer science form Novosibirsk State Technical University, Russian in 2002 and 2004, respectively.He joined doctor course at the University of Ulsan, Korea. He was awarded the Ph.D in 2010 from the same university. After conducting research as a postdoc fellow at the Ulsan University, he joined Korea University of Technology and Education as an assistant professor in 2011.
\end{IEEEbiography}




\end{document}